\documentclass[prl,twocolumn,superscriptaddress,showpacs]{revtex4}

\usepackage{epsfig}
\usepackage{amsmath,graphicx}
\usepackage{ae}
\usepackage{aecompl}

\begin{document}

\title{Copolymer adsorption kinetics at a selective liquid-liquid interface:\\
Scaling theory and computer experiment}
\author{A. Corsi}
\affiliation{Max - Planck - Institute for Polymer Research -
  Ackermannweg  10, 55128 Mainz, Germany}
 \author{A. Milchev}
\affiliation{Max - Planck - Institute for Polymer Research -
  Ackermannweg  10, 55128 Mainz, Germany}
\affiliation{Institute for Physical Chemistry, Bulgarian Academy of
  Sciences, 1113 Sofia, Bulgaria}
\author{V.G. Rostiashvili} 
\affiliation{Max - Planck - Institute for Polymer Research -
  Ackermannweg  10, 55128 Mainz, Germany}
\author{T.A. Vilgis} 
\affiliation {Max - Planck - Institute for Polymer Research -
  Ackermannweg  10, 55128 Mainz, Germany}

\pacs{36.20.-r, 68.05.-n, 07.05.Tp}
\date{\today}

\begin{abstract}
We consider the adsorption kinetics of a regular block-copolymer of total length $N$ and 
block size $M$ at a selective liquid-liquid interface in the limit of strong localization.
We propose a simple analytic theory based on scaling considerations which describes
the relaxation of the initial coil into a flat-shaped layer. The characteristic times
for attaining equilibrium values of the gyration radius components
perpendicular and parallel to the interface are predicted to scale with chain length $N$ 
and block length $M$ as $\tau_{\perp} \propto M^{1+2\nu}$ (here $\nu\approx 0.6$ is the 
Flory exponent) and as $\tau_{\parallel} \propto N^2$, although initially the rate of coil 
flattening is expected to decrease with block size as $\propto M^{-1}$. 
Since typically $N\gg M$ for multiblock copolymers, our results suggest that the flattening 
dynamics proceeds faster perpendicular rather than parallel to the interface.
We also demonstrate that these scaling predictions agree well with the results of 
extensive Monte Carlo simulations of the localization dynamics.
 
\end{abstract}

\maketitle

The behavior of hydrophobic - polar (HP) copolymers at a selective penetrable
interface (the interface which divides two immiscible liquids, like water and oil,
each of them being favored by one of the two types of monomers) is of great 
importance in the chemical physics of polymers. For strongly selective interfaces
the hydrophobic and polar blocks of a copolymer chain try to stay on different sides 
of the interface leading thus to a major reduction of the interfacial tension between 
the immiscible liquids or melts which has important technological applications, e.g. for 
compatilizers, thickeners or emulsifiers. Not surprisingly, during the last two decades 
the problem has gained a lot of attention from experiment \cite{Clifton,Rother,Wang,Omarjee}, 
theory \cite{Sommer1,Sommer2,Garel,Joanny,Denesyuk} as well as from computer experiment 
\cite{Balasz,Israels,Sommer3,Lyats,Chen}. While in earlier studies attention has been mostly
focused on diblock copolymers \cite{Rother,Omarjee} due to their relatively simple structure, 
the scientific interest shifted later to {\em random} HP-copolymers at penetrable
interfaces \cite{Sommer2,Garel,Joanny,Denesyuk,Chen}. Until recently though, the
properties of regular multiblock copolymers, especially with emphasis to their 
{\em dependence on block length} $M$, have remained largely unexplored. 

In a recent study \cite{Corsi} we showed that the equilibrium properties (structure, 
diffusion coefficient, etc.) of a regular HP-copolymer 
at a selective liquid-liquid interface are well described in both regimes of weak and strong
localization by a scaling theory in terms of the total copolymer length $N$ and the block
size $M$. In particular, we demonstrated that: (i) the crossover selectivity decreases with 
growing block length as $\chi_c \sim M^{-(1+\nu)/2}$ and the crossover selectivity to the 
strong localization regime vanishes as $\chi_{\infty} \sim M^{-1}$, and (ii)
the size of the copolymer varies in the weak localization regime as
$R_{g\perp} \sim M^{-\nu(1+\nu)/(1-\nu)}$ and $R_{g\parallel} \sim M^{(\nu_2 - \nu)
(1+\nu)/(1-\nu)}$, and as $R_{g\perp}\sim M^\nu,\quad R_{g\parallel} \sim M^{-(\nu_2-\nu)}$ 
in the case of strong localization (where $\nu_2 = 3/4$ is the Flory exponent in 
two dimensions). We shall use these results in the present communication in which we 
suggest a theory of the adsorption kinetics of a regular block copolymer at a liquid-liquid 
interface based on dynamical scaling arguments. To the best of our knowledge so far 
there have been no attempts to treat  this problem analytically or by means of computer 
simulation.
 
In our {\em dynamical scaling analysis} we consider a coarse-grained model of a multiblock 
copolymer consisting of $N$ repeat
units which is built up from a sequence of H- and P - blocks each of length $M$. 
For simplicity one may take the interface with negligible thickness as a flat plane which
separates the two selective immiscible solvents.  The energy gain of each repeat unit is 
thus $-\chi$, provided it stays in its preferred solvent, and the system is considered in 
the {\em strong localization} limit where $\chi >\chi_{\infty} \propto M^{-1}$
\cite{Corsi,Leclerc}. We neglect the effect of bulk diffusion on adsorption, and place the
 center of mass of an unperturbed coil  at time  $t = 0$ at the interface whereby the 
localization field is switched on. 
The initial coil will start then relaxing with time into a flat ("pancake") equilibrium 
configuration in-plane with the interface and the kinetics of relaxation will be determined by 
the sum of the various forces acting on the copolymer. 

In order to estimate the driving force of the flattening process, one may recall 
\cite{Corsi,Leclerc} that the effective attractive energy $\epsilon$ (per diblock) in the 
strong localization limit is $ \epsilon \approx \chi M $. In this case  the diblock (i.e.
a segment consisting of one H- and one P-block) plays the role of a blob and the overall 
attractive free energy $ F_{\rm attr} \propto \epsilon {\cal N} \propto \chi N$ where 
${\cal N}\simeq N/M $ is the total number of blobs. Thus  the effective driving force
perpendicular to the interface is $f_{\rm attr}^{\perp} \approx - 
\chi_{\infty}N/R_{\perp}$ where $R_{\perp}$ denotes the perpendicular component 
of the radius of gyration. This force is opposed by a force of confinement due to the 
deformation of the self-avoiding chain into a layer of thickness $ R_{\perp} $. The 
corresponding free energy of deformation is simply estimated as $ F_{\rm conf} 
\sim N(b/R_{\perp})^{1/\nu}$ where $b$ is the Kuhn segment size \cite{Gennes}.
For the respective force then one gets $f_{\rm conf}^{\perp} \simeq Nb^{1/\nu}/
R_{\perp}^{1/\nu + 1}$.

The equation of motion for $R_{\perp}$ follows from the condition that the friction force 
which the chain experiences during the motion in the direction perpendicular to the interface is balanced by the 
sum of $f_{\rm attr}^{\perp}$ and  $f_{\rm conf}^{\perp}$. In the case of Rouse dynamics, 
each chain segment experiences independent Stokes friction so that the resulting equation 
of motion has the form
\begin{eqnarray}
\zeta_{0} N \frac{d R_{\perp}}{d t}  = - \frac{\chi_{\infty}N}{R_{\perp}} + \frac{N b^{1/\nu}}{R_{\perp}^{1/\nu + 1}} \quad,
\label{Eq_motion_perp}
\end{eqnarray}
where $\zeta_{0}$ is the friction coefficient per segment.

During the flattening process, the chain spreads parallel to the interface due to
the excluded volume interaction. Within the Flory mean - field arguments, the corresponding 
free energy $F_{\rm ev}  \simeq v N^2 / (R_{\parallel}^2 R_{\perp})$ where $v$ is the second 
virial coefficient and $R_{\parallel}$ is the gyration radius component parallel to
the interface. 
The corresponding driving force is $f_{\rm ev}^{\parallel} = v N^2 / 
(R_{\parallel}^3 R_{\perp})$. This term is counterbalanced by the elastic force of chain
deformation, i.e. by $f_{\rm def}^{\parallel} =  - R_{\parallel}/(b N)$.

Taking into account the balance of forces in the parallel direction, the equation of motion 
for $R_{\parallel}$ takes then the following form
\begin{eqnarray}
\zeta_{0} N \frac{d R_{\parallel}}{d t}  = \frac{v N^2}{R_{\parallel}^3 R_{\perp}}  - \frac{R_{\parallel}}{b N} \quad.
\label{Eq_motion_parallel}
\end{eqnarray}
Evidently, the excluded volume interactions provide a coupling between the relaxation perpendicular and parallel to the interface.
Thus Eqs. (\ref{Eq_motion_perp}) and (\ref{Eq_motion_parallel}) describe the relaxation 
kinetics of a multiblock copolymer conformation at a selective liquid-liquid interface. 
One may readily verify that the equilibrium solutions which follow from these equations 
are
\begin{eqnarray}
\label{Equilibrium_perp}
R_{\perp}^{\rm eq} \simeq b M^{\nu}, 
\end{eqnarray}
and 
\begin{eqnarray}
\label{Equilibrium_paral}
R_{\parallel}^{\rm eq} \simeq (v b)^{1/4} M^{-\nu/4} N^{\nu_{2}}.
\end{eqnarray}
This coincides with the equilibrium expressions for $R_{\perp}$ and $R_{\parallel}$ 
derived earlier from purely scaling consideration \cite{Corsi}. The only difference is in 
the power of the $M$ - dependence in eq.(\ref{Equilibrium_paral}) which looks like 
$M^{-\nu/4}$ instead of $M^{-(\nu_{2} - \nu)}$ in ref. \cite{Corsi} albeit numerically 
the values of both exponents coincide: $\nu/4 \simeq (\nu_{2} - \nu) \simeq  0.15$.

For the full solution of the equations of motion it is convenient to rescale the variables 
as $x \equiv R_{\perp}/(b M^{\nu}),\quad y \equiv R_{\parallel}^{4} M^{\nu}/(v b N^3)$ 
so that eqs.(\ref{Eq_motion_perp}) and (\ref{Eq_motion_parallel}) can be written in the 
dimensionless form 
\begin{eqnarray}
\tau_{\perp} \: \frac{d x}{d t} &=& \frac{1}{x^{1/\nu + 1}} - \frac{1}{x} \label{Eq_motion_x}\\
\tau_{\parallel} \: \frac{d y}{d t} &=& \frac{1}{x} - y \quad.
\label{Eq_motion_y}
\end{eqnarray}

The characteristic times for relaxation perpendicular and parallel to the interface in 
eqs. (\ref{Eq_motion_x})-(\ref{Eq_motion_y}) should then scale as $ \tau_{\perp} 
\simeq \zeta_0 b^2 M^{1 + 2\nu} $, and $\tau_{\parallel} \simeq \zeta_0 b^2 N^2 $.

Eq. (\ref{Eq_motion_x}) can be solved exactly:
\begin{eqnarray}
\frac{x^2(t)}{2}\left[ 1 - F(2 \nu,1;1 + 2 \nu; x^{1/\nu}(t))\right] \nonumber \\
- \frac{x^2(0)}{2}
\left[ 1 - F(2 \nu,1;1 + 2 \nu; x^{1/\nu}(0))\right] = - \frac{t}{\tau_{\perp}} \quad,
\label{Solution}
\end{eqnarray}
where $ F(\alpha, \beta; \gamma; z) $ is the hypergeometric function and $ x(0) = 
R_{\perp}(0)/b M^{\nu} $ is the initial value. In the {\em early stage} of relaxation (i.e. at 
$t \ll \tau_{\perp}$) one has $x \gg 1$ and the solution, eq.(\ref{Solution}), with 
$ R_{\perp}(0) \simeq b N^{\nu} $ reduces to
\begin{eqnarray}
R_{\perp}^{2}(t)  =  R_{\perp}^{2}(0) - \frac{t}{\zeta_{0} M} \quad,
\label{Solution_limit11}
\end{eqnarray}
so that the perpendicular collapse of the chain should last proportionally to the block 
length $M$.
In the opposite limit of {\em late stage} kinetics, $t \approx \tau_{\perp}$ and 
$x \geq 1$, that is, close to equilibrium, the relaxation of $R_{g}(t)$ is essentially 
exponential with $\tau_{\perp}\propto M^{2.2}$.
\begin{eqnarray}
\frac{R_{\perp}(t)}{ R_{\perp}^{\rm eq} } \simeq 1 +\exp(-\frac{t}{\tau_{\perp}})\quad.
\label{Solution_limit2}
\end{eqnarray}

Moreover, as typically $ N \gg M $, one expects that $\tau_{\parallel} \gg 
\tau_{\perp}$, i.e. the chain coil collapses first in the perpendicular direction to its 
equilibrium value and after that slowly extends in the parallel direction. In this case one
can use $ x(t) \approx x^{\rm eq} =1 $ in eq.(\ref{Eq_motion_y}) and derive the resulting 
solution for the parallel component of $R_g$ as
\begin{eqnarray}
y(t) - 1 = \left[y(0) - 1\right]\exp(-t/\tau_{\parallel}) \quad,
\label{Solution_y}
\end{eqnarray}
where the initial value $ y(0) < 1 $. 
\begin{figure}[htb] \begin{minipage}[t]{0.5\textwidth}%
\vspace{1mm}
  \hspace*{-9mm}
  \setlength{\epsfxsize}{0.9\textwidth}
  \epsffile{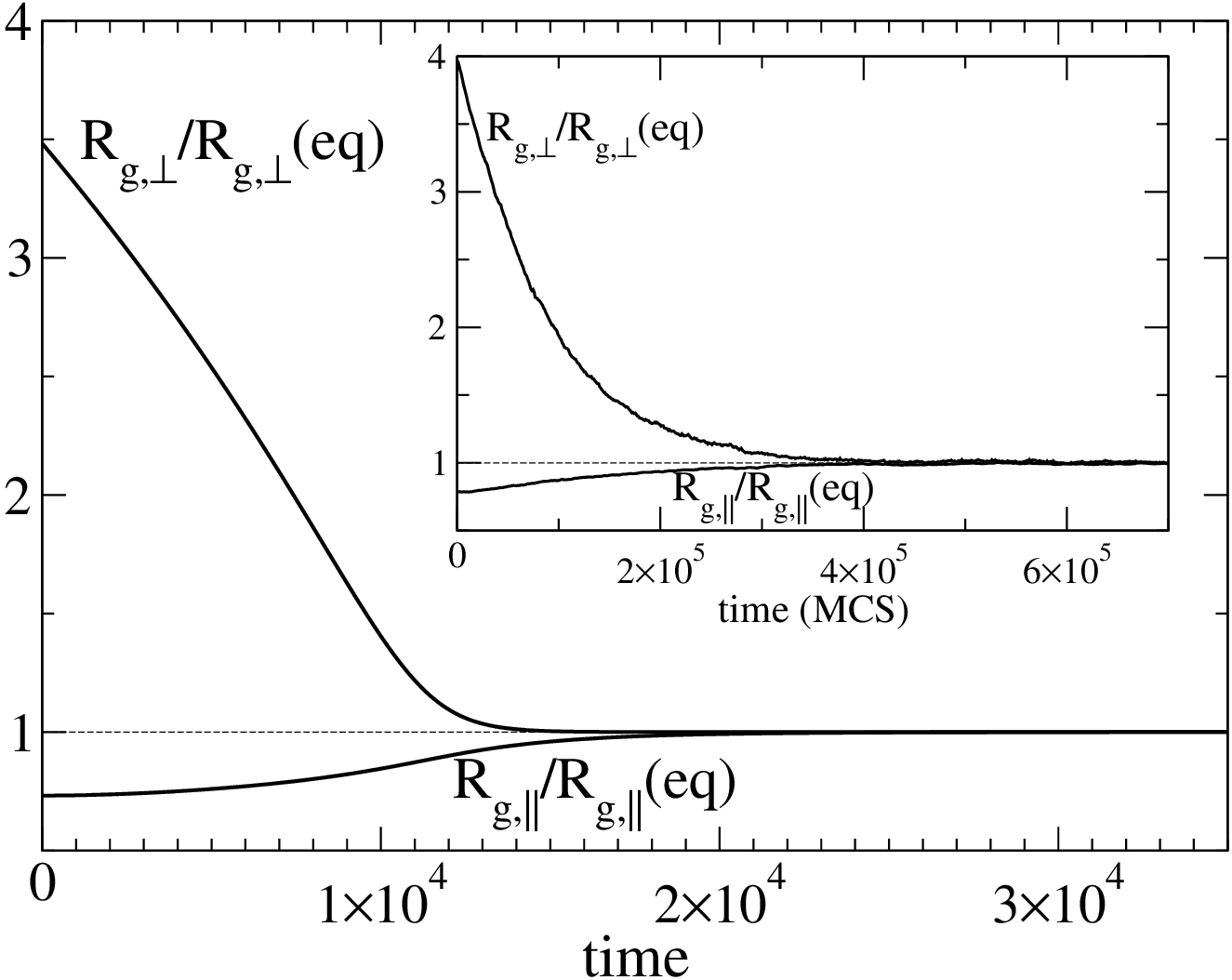}
\caption{\label{fig1}
Relaxation of $R_\perp$ and $R_\parallel$ with time, following from the numerical solution of 
eqs. (\ref{Eq_motion_x})-(\ref{Eq_motion_y}) for a copolymer of length $N = 128$ and
block size $M = 16$. Both components are normalized by their respective equilibrium 
values. The inset shows the result for the same quantities (and the same set of 
parameters) as obtained from the MC simulations.}
\end{minipage}
\end{figure}
As one may see in Fig. \ref{fig1}, the time evolution of $R_\perp$ and $R_\parallel$
which follows from the solution of the system of differential equations (\ref{Eq_motion_x}) - 
(\ref{Eq_motion_y}), resembles qualitatively rather well the simulation data even though
one should bear in mind that the time axis for the former is in arbitrary units (scaling
always holds up to a prefactor) whereas in the simulation time is measured in MC steps (MCS)
per monomer (i.e., after all monomers have been allowed to perform a move at random). 
The relaxation of $R_\parallel$ looks somewhat slower and, as our data for longer 
chains show, this effect becomes much more pronounced with growing number of blocks.

For the {\em Monte Carlo simulations} we use an off-lattice bead-spring model that has 
been employed 
previously for simulations of polymers both in the bulk and near confining surfaces \cite{KBAM}.
Recently it was applied by us to the study of the static properties of block copolymers at 
a selective interface \cite{Corsi}, therefore we describe here the salient features only. 

Each polymer chain contains $N$ effective monomers connected by anharmonic springs 
described by the finitely extendible nonlinear elastic (FENE) potential,
$U_{FENE} =-\frac{K}{2} R^2 \ln \big[1-\frac{(\ell - \ell_0)^2}{R^2} \big]$.
Here $\ell$ is the length of an effective bond, which can vary in between $\ell_{min}< \ell
<\ell_{max}$, with $\ell_{min}=0.4$, $l_{max}=1$ being the unit of length, and has the
equilibrium value $l_0=0.7$, while $R=\ell_{max}-\ell_0=\ell_0-\ell_{min}=0.3$, and the spring
constant $K$ is taken as $K/k_BT=40$. The nonbonded interactions between the effective 
monomers are described by the Morse potential, 
$U_M=\epsilon_M \{\exp[-2 \alpha(r-r_{min})]-2 \exp [-\alpha(r-r_{min})]\} \,$
where $r$ is the distance between the beads, and the parameters are chosen as $r_{min}=0.8$, 
$\epsilon_M=1$, and $\alpha=24$. Owing to the large value of the latter constant, $U_M(r)$ 
decays to zero very rapidly for $r>r_{min}$, and is completely negligible for distances larger 
than unity. This choice of parameters is useful from a computational point of view, since it 
allows the use of a very efficient link-cell algorithm \cite{KBAM}.  From a physical point
of view, the interactions $U_{FENE}$ and $U_M$ 
make sense when one interprets the effective bond as
a Kuhn segment, comprising a number of chemical monomers along the chain, and thus the 
length unit $\ell_{max}=1$ corresponds physically rather to 1 nm than to the length of a 
covalent $C-C$ bond (which would only be about $1.5 \text{\r{A}}$). Since in the present study 
we are concerned with the localization of a copolymer at {\em good solvent} conditions, in 
$U_M(r)$ we retain the repulsive branch of the Morse potential only by setting
$U_M(r) = 0 \quad \mbox{for} \quad r > r_{min}$ and shifting $U_M(r)$ up by $\epsilon_M$.

The interface potential is taken simply as a step function with amplitude $\chi$,
$U_{int}(n,z)=\{-\sigma(n)\chi/2, z > 0;\quad \sigma(n)\chi/2, z \le 0 \}$
where the interface plane is fixed at $z = 0$, and $\sigma(n) = \pm 1$
denotes a "spin" variable which distinguishes between P- and H- monomers. 
In studying the flattening dynamics of these chains, we always start with a 
configuration which has been equilibrated in a solvent that is good for both types 
of monomers. At time $t=0$, the interface is switched on (so that it goes through 
\begin{figure}[htb] \begin{minipage}[t]{0.5\textwidth}%
\vspace{5mm}
  \hspace*{-9mm}
  \setlength{\epsfxsize}{0.9\textwidth}
  \epsffile{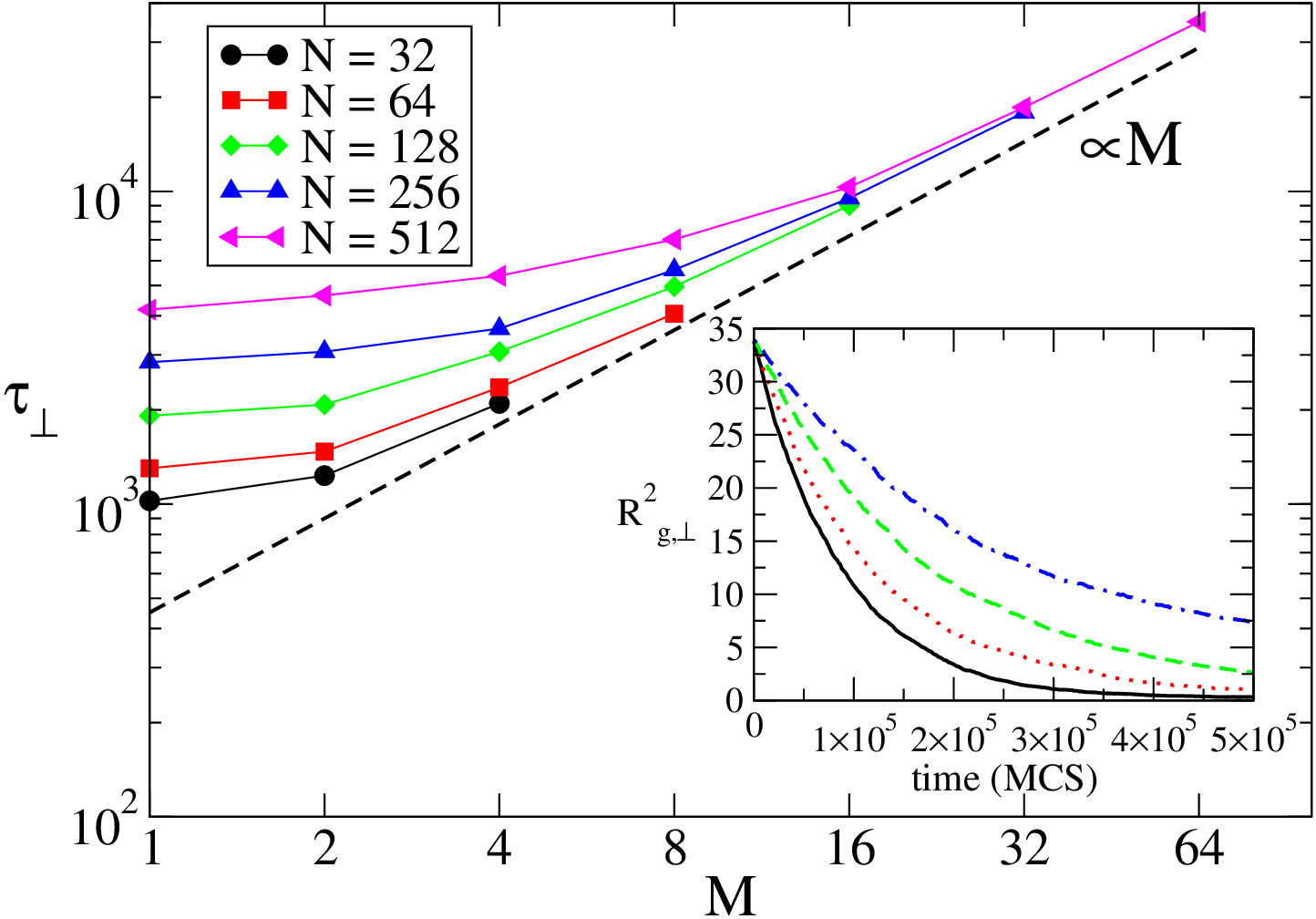}
\end{minipage}
\begin{minipage}[b]{0.5\textwidth}%
\caption{\label{fig2}
Variation of the characteristic time for the initial relaxation of $R_\perp$ with 
block length $M$ for chains of length $32 \le N \le 512$. The theoretical prediction 
is shown by a dashed line. The inset shows the typical behavior of $R_\perp (t)$ 
in the case of $N = 256$, and $M = 2$ (solid line), $4$ (dotted line), $8$ (dashed 
line) and $16$ (dot-dash line).}
\end{minipage}
\end{figure}
the center of mass of the chain) and the flattening dynamics is then followed until 
the chain reaches its new equilibrium configuration at the interface. Since we are 
interested in the behavior of these chains in the limit of strong localization, we 
have chosen $\chi=2\chi_{c}$ where $\chi_{c}$ is the crossover selectivity as obtained 
in Ref. \cite{Corsi}. We use periodic boundary conditions in the plane
of the interface while there are rigid walls in the z - direction, where the simulation box
extends from $z = -32$ to $z = 32$. Typically we studied chains with lengths $32 \le N \le 512$ 
and block lengths $1\le M \le N/8$ whereby all measurements have been
averaged over $1024$ different and independent equilibrated starting configurations.

\begin{figure}[htb] \begin{minipage}[t]{0.5\textwidth}%
\vspace{6mm}
  \hspace*{-9mm}
  \setlength{\epsfxsize}{0.9\textwidth}
  \epsffile{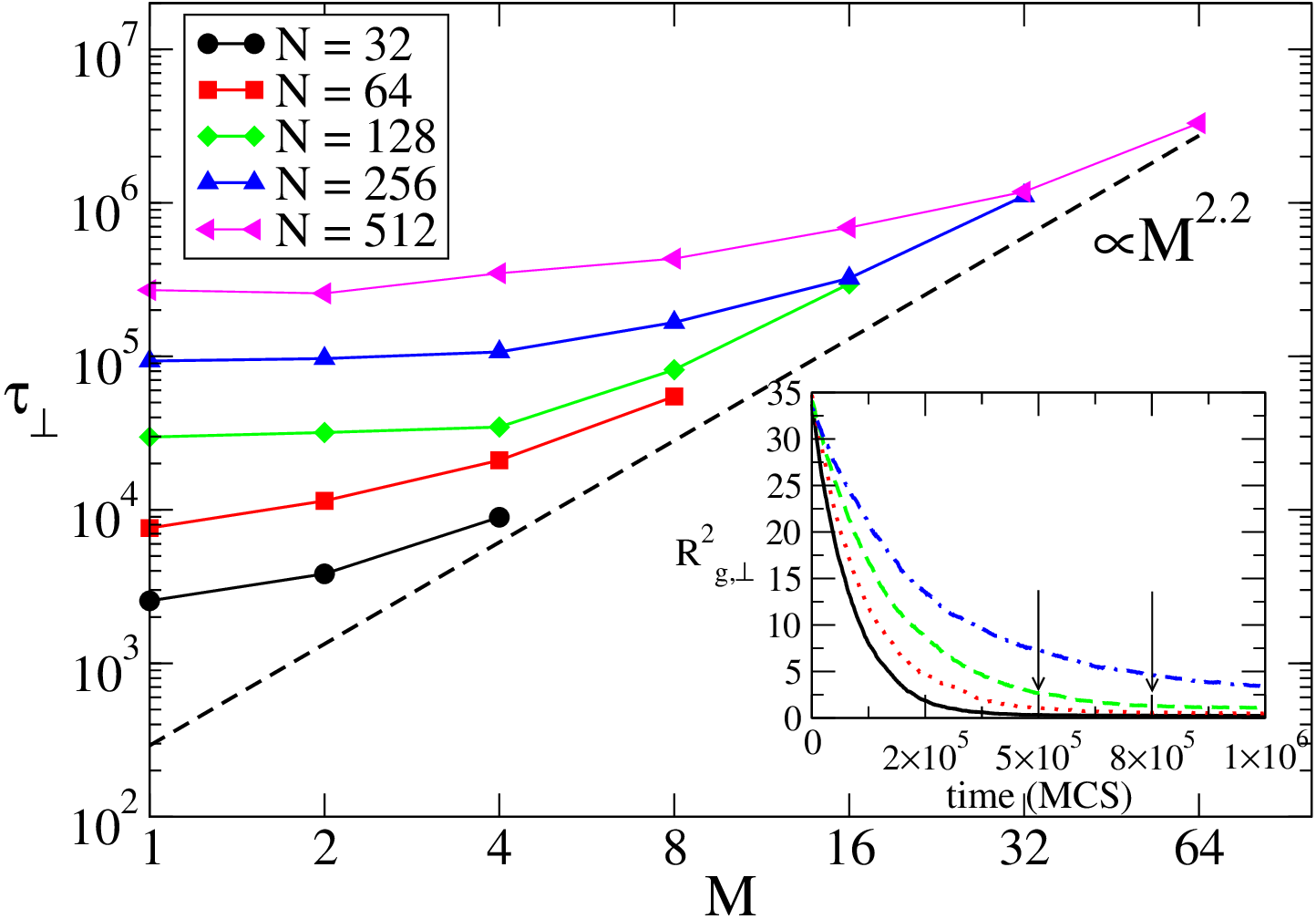}
\caption{\label{fig3}
Variation of $\tau_\perp$
with block length $M$ for chains of length $32 \le N \le 512$. The slope of the dashed
line is $\approx 2.2$, according to  eq. (\ref{Solution_limit2}). 
The inset shows the typical behavior of $R_\perp (t)$ in the case of $N = 256$,
and $M = 2$ (solid line), $4$ (dotted line), $8$ (dashed line) and $16$ (dot-dash 
line). Arrows denote the time interval where $\tau_\perp$ has been determined by 
regression in the case of $M = 8$.}
\end{minipage}
\end{figure}

From Fig. \ref{fig2} one may readily verify that the initial collapse of $R_\perp
(t\ll \tau_\perp )$ with time closely follows the predicted rate $\propto M^{-1}$ 
according to eq. (\ref{Solution_limit11}) in the limit of $N, M\gg 1$. 
Moreover, for sufficiently large block size $M$ the $N$-dependence of this initial rate
disappears so that asymptotically this initial perpendicular collapse is governed
by the length $M$ only. In contrast, during the late stages of localization at the
interface one observes on Fig. \ref{fig3} the expected, cf. eq. (\ref{Solution_limit2}), 
characteristic scaling of $\tau_\perp \propto M^{1+2\nu}$ which progressively improves
as the asymptotic limit is approached. Note that the fits of $\tau_\perp$ are taken 
only {\em after} the initially unperturbed coil has sufficiently relaxed, $R_\perp (t) /
R_\perp (0) \le 1/e$. As expected, the initial dependence of  $\tau_\perp$ on $N$
is seen to vanish as $M$ and $N$ become sufficiently large, in agreement with eq. 
(\ref{Solution_limit2}).

\begin{figure}[htb] \begin{minipage}[t]{0.5\textwidth}%
\vspace{7mm}
  \hspace*{-9mm}
  \setlength{\epsfxsize}{0.9\textwidth}
  \epsffile{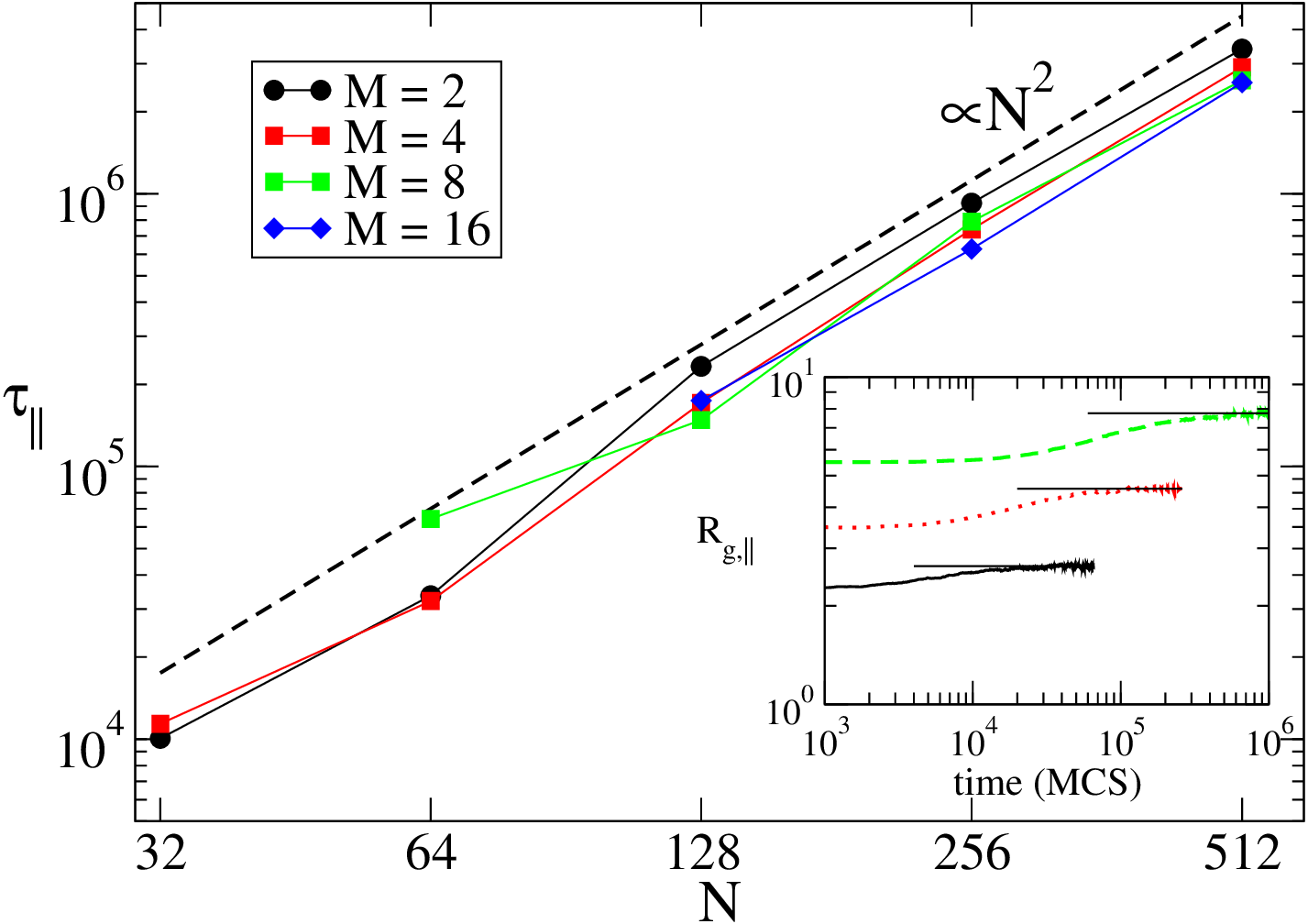}
\caption{\label{fig4}
Variation of $\tau_\parallel$ with chain length $N$ for blocks of size $M = 2, 4, 8$ and $16$.
The inset shows the typical behavior of $R_\parallel (t)$ in the case of $M = 4$,
and chain lengths $N = 32$ (solid line), $64$ (dotted line), and $128$ (dashed line).
The respecive equilibrium values are indicated by horizontal lines.}
\end{minipage}
\end{figure}

Eventually, in Fig. \ref{fig4} we display the measured scaling of the characteristic
time for spreading along the interface, $\tau_\parallel$, with block size $M$ for
different chain lengths $32\le N\le 512$. Apart from some scatter of data for too small
${\cal N} = N/M$, one recovers nicely the relationship $\tau_\parallel \propto N^2$
whereby, once again, the initial $M$-dependence gradually diminishes as ${\cal N} \gg 1$.

In summary, it appears that the simple scaling theory of copolymer adsorption on a 
selective liquid-liquid interface captures well the most salient features of the problem
as a comparison with extensive MC simulation data demonstrates. 
In this letter we focus our consideration on the case of Rouse dynamics whereby hydrodynamic 
effects are neglected. Therefore the present approach should be appropriate to the case of 
an interface between immiscible polymer melts. 
The case of the so called Zimm dynamics which accounts for
hydrodynamic effects will be discussed in an extended paper.

{\em Acknowledgments} AM acknowledges the support and hospitality of the Max-Planck Institute 
for Polymer Research in Mainz during this study. This research has been supported by the
Sonderforschungsbereich (SFB 625).

\end{document}